\documentstyle[12pt]{article}
\def\be{\begin{equation}}
\def\ee{\end{equation}}
\def\bea{\begin{eqnarray}}
\def\eea{\end{eqnarray}}
\def\a{\alpha}
\def\b{\beta}

\def\e{\epsilon}
\def\l{\lambda}

\author{Hans - J\"urgen Schmidt}

\title{On Ellis' programme within fourth order gravity}

\date{}
\begin{document}
\maketitle

\centerline{Universit\"at Potsdam, Institut f\"ur Mathematik, Am
Neuen Palais 10} 
 \centerline{D-14469~Potsdam, Germany,  E-mail:
 hjschmi@rz.uni-potsdam.de}

\begin{abstract}
For the non-tachyonic curvature squared action we show that the 
expanding Bianchi-type I models tend to the dust-filled 
Einstein-de Sitter model for $t$ tending to infinity if 
the metric is averaged over the typical oscillation period. 
Applying a conformal equivalence between curvature squared 
action and a minimally coupled scalar field (which holds for all 
dimensions $>$ 2) the problem is solved by discussing 
a massive scalar field in an anisotropic cosmological model.

\bigskip

\noindent 
F\"ur das tachyonenfreie Wirkungsintegral mit 
Kr\"ummungsquadraten zeigen wir, da\ss \  die expandierenden 
Bianchityp I-Modelle f\"ur $t \to \infty$  gegen das 
staubgef\"ullte Einstein-de Sitter-Modell konvergieren, 
falls die Metrik \"uber die typische Oszillationsperiode 
gemittelt wird. Wendet man die Konform\"aquivalenz der 
Kr\"ummungsquadratwirkung zu einem minimal 
gekoppelten Skalarfeld an  (die f\"ur jede Dimension $>$ 2
 g\"ultig ist), kann man das Problem durch Betrachtung 
eines massebehafteten Skalarfeldes 
in einem anisotropen kosmologischen Modell l\"osen.
\end{abstract}

AAA classfication: 162

\section{Introduction}

ELLIS (1984) has asked on
 which length scale the Einstein field equation is valid. 
On laboratory or on cosmic distances? Here 
we extend this question to cover microscopic 
scales, too. This is a real problem as one knows:

\noindent
1.   An averaging procedure 
does not commute with a non-linear 
differential operator as the Einstein tensor is.

\noindent
2.   Einstein's theory is well tested at large distances $\gg$ 1 cm.

\noindent 
3. The ultraviolet divergencies of Einstein gravity 
can be removed by adding curvature squared terms 
to the action, WEINBERG (1979); this is a microscopic 
phenomenon and forces  to prefer a curvature squared action 
already on the level of classical field theory, as we are concerned 
here.

\bigskip

Now, we propose a synthesis of 1., 2., and 3. as follows: 
microscopically, we take
\be
     L_{\rm g} = (R/2 - l^2 R^2)/8\pi G 
\ee
(the non-tachyonic case). By an averaging procedure 
we get Einstein gravity 
on large scales $ \gg l \approx   10^{-28}$  cm. 
The curvature 
squared contribution represents effectively dust in 
the asymptotic 
region $t \to \infty$.  For spatially flat Friedmann 
models  this was already proven in M\"ULLER
 and SCHMIDT (1985), here we generalize to models 
with less 
symmetry. In the present paper we consider only the 
vacuum case
$$
\delta L_{\rm g} \sqrt{- \det g_{ij}}
 \delta g^{kl}   = 0 \,   ,
$$
 and  therefore, we interpret the effectively obtained dust  
as invisible 
gravitating matter necessary to get a spatially flat universe.

\bigskip

We conjecture that additional matter contributions (usual 
dust plus radiation) do not alter the result qualitatively. 
The general expectation for Lagrangian (1) 
is the following: starting at die Planck era
$$
R_{ijkl}R^{ijkl}
\approx  10^{131} \, {\rm  cm}^{-4} \, ,
$$
the $R^2$-term is dominant,  the inflationary de 
Sitter phase is an attractor; its appearance becomes very 
probable (SCHMIDT 1986, STAROBINSKY and SCHMIDT 1987); 
the $R$-term in (1) yields a 
parametric decay of the value of $R$ and 
one turns over to the
 region $t \to \infty$ where $R^2$  gives just dust in the mean.

\bigskip

In CARFORA and MARZUORI (1984) 
another approach to Ellis' programme was initiated: a 
smoothing out of die spatially closed 3-geometry in the
  direction of die spatial Ricci-tensor 
leads to different values of mean mass density 
before and after the smoothing out procedure.

\section{The massive scalar field 
in a Bianchi-type I comological model}

Consider  
a minimally coupled scalar 
field $\phi$  in a potential $V(\phi)$  ($8\pi G = 1$)
$$
 L=R/2-   \frac{1}{2} \phi_{;i} \phi^{;i}     + V(\phi)    \, .
$$
We suppose  $V(\phi)$  to be a $C^3$-function and 
$V(0) = 0$  a local quadratic minimum of the potential $V$,
 i.e., $V'(0) = 0, V"(0) =m^2$, $m > 0$. To describe the
 asymptotic behaviour $\phi \to  0$ as $t \to \infty$
 it suffices to use 
$V(\phi) =  m^2 \phi^2/2 $ because the higher
order terms 
do not affect it.   (This statement can be proved  as follows: 
For each $\e > 0$, $\e < m$, there exists a 
$\phi_0 > 0$ such that for all $\phi $ with $\vert \phi \vert < 
\phi_0 $ it holds
$$
(m - \e)^2 \phi^2 \le 2V(\phi) \le  (m + \e)^2 \phi^2 \, .
 $$
And then all further development is enclosed by inequalities with 
$m \pm \e$, $\e \to  0$.)

\bigskip

Here, we concentrate on a 
cosmological model of Bianchi-type I. It can be written as
\be
ds^2=   dt^2 - e^{2\a } [ e^{2(s+\sqrt{3}r)} dx^2
 + e^{2(s- \sqrt{3}r)} dy^2 + e^{-4s} dz^2]    \, .
\ee
$\phi$, $\a$, $r$, $s$
 depend on $t$ only. $\dot \a = d\a /dt = h$
 is the Hubble parameter and $\eta = u/h$, 
$ u =   (\dot r^2 + \dot s^2)^{1/2}$  the anisotropy parameters. 
It holds $0 \le \eta \le 1$, $\eta = 0$
 represents the isotropic model, and $\eta = 1$
 gives $\phi \equiv 0$, therefore, we consider only 
the case $0 < \eta < 1$ in the following. $h = 0$
 is possible for Minkowski space-time only, and 
the time arrow is defined by $h> 0$,
 i.e., we restrict to expanding solutions. In LUKASH, SCHMIDT (1987) 
it was shown that all such solutions tend to Minkowski 
spacetime for $t \to \infty$.
 More detailed: At $t = 0$, let $\a = r = s = 0$ by a coordinate
transformation. 
Then for each prescribed quadruple
 $ (\dot r, \, \dot s, \, \phi, \, \dot \phi)$
  at $t = 0$  the integration 
of the relevant system (We put the mass $m = 1$.)
\bea
     (\phi^2  + \dot \phi^2 + u^2)^{1/2} = h \\
\ddot \phi  +  3h \dot \phi + \phi = 0       \\
\dot r = C_r e^{-3\a} \, , \qquad    \dot s  = C_s e^{-3\a}
\eea
($C_r$, $C_s$  are constants,)  up to $t \to \infty$  is 
possible with $r$, $s$, $\phi$, $\dot \phi$, $h$, $\dot h$
  tending 
to zero in that limit. Now, we consider 
this limit in more details. Up to now the 
following is known: For the isotropic 
models $u \equiv 0$, $e^\a = a$ one knows that the 
asymptotic behaviour is given by oscillations around 
$a \sim  t^{2/3}$  i.e., we get the Einstein-de Sitter model 
in the mean, and the effective equation of 
state is that of dust, cf. e.g. STAROBINSKY (1978). 
In GOTTL\"OBER (1987) this
 is generalized to a special class
 of inhomogeneous, nearly isotropic models.
After averaging over
 space and oscillation period one gets as effective equation of state
\be
p \sim \rho/a^2
\ee
i.e., also dust in the limit $t \to \infty$. 

\bigskip

The question for the Bianchi-type I model 
is now: does one get the energy-momentum tensor 
of an ideal fluid in the mean, 
or will there appear a strongly anisotropic pressure? 
The first question to be  answered is about 
the averaging procedure:
From eq. (4) and the fact that $h \to  0$  as $t \to \infty$  one 
gets a fixed oscillation period: let $t = t_n$ be  the $n$-th
 local extremum of $\phi(t)$, then
$$
\lim_{n \to \infty} t_{n+1} - t_n = \pi   \, .
$$
Therefore, we average the metric about this period of time. 
Let us denote $r(t_n)$  by $r_n$, \dots \, 
 For a fixed finite time it is ambiguous how to perform this average, 
but in the limit $t  \to \infty$
 one can describe 
the system by adiabatic 
(or parametric) deviations from 
pure $\phi \sim   \cos t$-oscillations, and 
the averaging procedure consists of 
constructing   monotoneous smooth curves $\bar \a (t)$
 fulfilling $\bar \a(t_n) = \a_n$, \dots  and the effective 
energy-momentum tensor is obtained 
from eq. (1) 
with $\a = \bar \a (t)$, \dots  using the Einstein equation.

\bigskip

Derivating eq. (3) one gets with (4)
\be
0 \ge \dot  h = 3\phi^2 - 3h^2 \ge  -3h^2 
\ee
and therefore,
$$
h_n \ge h_{n+1} > h_n  - 3\pi h_n^2 \, .
$$
Let 
$$
h   \ge \frac{1}{3\pi (n+n_0)} 
$$
be valid for one value $n = n_1$  then, by induction, 
 this holds true also for all larger values $n$, i.e.,
\be
h(t) \ge \frac{1}{3(t+t_0)}
\ee
for some $t_0$,  let $t_0 = 0$ subsequently. 
Derivating (5) we get with  eq. (3)
\be
\dot \eta 
= - 3 \eta (1- \eta^2 ) h \cdot 
\frac{\phi^2}{\phi^2 + \dot \phi^2} \, .
\ee
The last factor can be substituted by its mean value, $= 1/2$.
 From eq. (8) one
 can see that for initial conditions $0 < \eta < 1$
as we met here, eq. (9) leads to $ \eta \to 0$  as $t \to \infty$.
 Knowing this, we can perform a stronger estimate for $h$, 
because in the mean,
 $\dot h = -3h^2/2$, 
i.e., in the leading order we have $h(t) = 2/3t$  and, using (9)
$$
\dot \eta =- \eta/t \, , \qquad    \eta = \eta_0/t \, .
$$
Remember: the inflationary era diminishes $\eta$  exponentially, but 
here we consider 
only the asymptotic region. In sum  we get for metric (2)
\be
e^\a  = t^{2/3}\, , \qquad r=-  C_r/t \,  ,
 \quad    s = -C_s/t \,   . 
\ee
Inserting (10) into (2) we get, via the 
Einstein equation, a diagonal energy-momentum tensor 
with $\rho  = T_{00} = 4/3t^2$  and pressure of the 
order $ (C_r^2 + C_s^2)/t^4$. The effective equation of 
state is $p \sim \rho/t^2 \sim \rho/a^3$, i.e., also dust in the 
limit $t \to \infty$  as eq. (6). From the details of the 
paper  of  LUKASH and SCHMIDT (1987) one can 
see that for almost all models the end of the 
inflationary stage 
is just the beginning of the Einstein-de Sitter stage.

\bigskip

A similar result occurs if not the metric, 
but $\rho$ and $p$ will be averaged, cf.  GOTTL\"OBER (1984).

\bigskip

Result: For the minimally coupled 
scalar field in a potential $V$ which has  a  single 
quadratic minimum at $V = 0$  all expanding Bianchi-type I 
cosmological solutions tend to the Einstein-de Sitter model 
for $t \to \infty$  if 
the metric is averaged over the oscillation period.

\section{The generalized equivalence}

Some types of  a conformal equivalence
 theorem between fourth order gravity 
and minimally coupled scalar fields are obtained 
in SCHMIDT (1987, 1988) and ref. cited therein. 
This theorem was independently obtained 
by FERRARIS (1986)  (cited from JAKUBIEC 
and KIJOWSKI (1987) and GOENNER (1987)) 
and STAROBINSKY (1988), too. All of them are 
restricted to 4-dimensional space-times. On the other hand, 
both $R^2$-terms and scalar fields have been discussed 
for higher-dimensional space-times, recently, 
cf.  e.g. ISHIHARA (1986). Therefore, it is worth 
mentioning that this conformal equivalence 
theorem can be formulated for arbitrary dimensions $n > 2$: Let
\be
\tilde{\cal L}
 = \tilde 
 R/2-   \frac{1}{2}\tilde g^{ij} \phi_{\vert i}
 \phi_{\vert j}     + V(\phi)    
\ee
and
\be
\tilde g_{ij}= e^{\l \phi} g_{ij} \, , \qquad \l =
\frac{  2}{ (n-1)(n-2)}
\ee
be the conformally transformed metric. Then 
the solutions of the variation of  (11) are transformed 
by (12) to the solutions of the variation of $ L = L(R)$, where
\be
 R = - 2e ^{\l \phi} \left(
\frac{nV}{n-2} + \mu \frac{dV}{d \phi}
\right) \, , \qquad 
\mu = \sqrt{\frac{n-1}{n-2}}
\ee
is supposed to  be locally (round $R = R_0$)  invertible as
\bea
 \phi=F(R) \, ,     
 \qquad F(R_0)=0 \, , \qquad F'(R_0) \ne 0   \nonumber \\
{\cal L}(R) = \frac{1}{2} R_0 + V(0) + 
 \frac{1}{2} \int_{R_0}^R e^{F(x)/\mu}  dx    \, .
\eea
The inverse direction is possible 
provided 
 $L'(R) L"(R) \ne 0$, 
cf.  SCHMIDT (1987, 1988) for details with $n = 4$.

\section{The fourth order gravity  model}

Now we come to the question posed  in the 
introduction: Let $L(R)$  be a $C^3$-function 
fulfilling $L(0) = 0$, $L'(0) L"(0) < 0$. Then we can write
\be
{ \cal L} (R) = \frac{R}{2 } + \b R^2 + O(R^3) \, , \qquad \b <0 \, .  
\ee
We consider the Bianchi-type I vacuum 
solutions which start in a neighbourhood of the
 Minkowski space-time and ask for the 
behaviour as $t   \to \infty $. Applying the
 equivalence theorem cited in sect. 3 we arrive at the
 models discussed in sect. 2, and this is applicable for
$\vert R \vert $  being small enough. The conformal factor 
depends on $t$ only, and therefore, the space of 
Bianchi-type I models
 will not be leaved, and we can formulate the 
following: In a neighbourhood of Minkowski space-time, all 
Bianchi-type I  models which represent a stationary point
 of the action (15), can be integrated 
up to $t   \to \infty $ 
or  $-  \infty $, let it be $ + \infty $. One singular solution 
is the Kasner solution and all other solutions undergo 
isotropization and have an averaged equation of state $p = 0$
 for $t   \to \infty $.

\section{Discussion}

ANDERSON (1986) discussed the 
possibility that curvature squared terms 
give an effective contribution to the 
energy density  of a Friedmann model. 
This could explain the discrepancy between the 
observed mean mass density   of about
 1/10 the critical one and the predicted 
(from inflationary cosmology) nearly the 
critical one. Here, we have shown that also 
for a large initial anisotropy the oscillating curvature
 squared contributions give just dust in the mean 
and not an equally large anisotropic pressure 
as one could have expected. The next 
step would be to look for a generalization 
of this fact to inhomogeneous cosmological models.

\section*{References}

\noindent 
ANDERSON, P.: 1986, Phys. Lett. B {\bf 169}, 31.

\noindent 
CARFORA, M. and MARZUOLI, A.: 1984, Phys. Rev. Lett. {\bf 53}, 2445.

\noindent 
ELLIS, G. F. R.: 1984, Ann. Rev. Astron. Astrophys. {\bf  22}, 157.

\noindent 
FERRARIS, M.: 1986, Atti del
 VI Convegno Nazionale di Relativita 
Generale e Fisica della Gravitazione, 
Editors: R. FABBRI, M. MODUGNO, Pitagora Editrice Bologna, p. 127.

\noindent 
GOENNER, H.: 1987, Proc. 11th Int. Conf. 
Gener. Rel. Grav. Ed.: M. MAC CALLUM, Cambridge Univ. Press, p. 262.

\noindent 
GOTTL\"OBER, S.: 1984, Astron. Nachr. {\bf 305}, 1.

\noindent 
GOTTL\"OBER, S.: l987, Astrophys. Space Sc. {\bf  132}, 191.

\noindent 
ISHIHARA, R.: l986, Phys. Lett. B {\bf 179}, 217.

\noindent 
JAKUBIEC, A., and KIJOWSKI, J.: 1987, preprint.

\noindent 
LUKASH, V. N., SCHMIDT, H.-J.: 1987, Preprint  
IKI Moscow, 1988, Astron. Nachr. {\bf 309}, 25.

\noindent 
M\"ULLER, V., SCHMIDT, H.-J.: 1985, Gen. Relativ. Grav. {\bf 17}, 769.

\noindent 
SCHMIDT, H.-J.: 
l986, thesis B, Berlin, GDR Academy of Sciences, unpublished.

\noindent 
SCHMIDT, H.-J.: l987, Astron. Nachr. {\bf  308}, 183.

\noindent 
SCHMIDT, H.-J.: 1988, PRE-ZIAP 87-06; Class. Quant. Grav. {\bf 5}, 233.

\noindent 
STAROBINSKY, A. A.: 1978, Pis'ma Astron. Zh. {\bf 4}, 155.

\noindent 
STAROBINSKY, A. A.: 1988, Proc. 4th Sem.
 Quantum Grav. Moscow, Eds. M. A. MARKOV, V. P. FROLOV, 
to appear. 

\noindent 
STAROBINSKY, A. A., 
SCHMIDT, H.-J.: 1987, Class. Quant. Grav. {\bf 4}, 695.

\noindent 
WEINBERG, S.: 1979, in HAWKING, S., ISRAEL, W. (Eds.) 
General Relativity, Cambridge Univ. Press. p. 790.

\bigskip

Received 1987 November 17

\medskip

\noindent 
{\small {This is a  reprint from Astronomische Nachrichten, 
done with the kind permission of the copyright owner, only some 
obvious
 misprints have been cancelled;    
 Astron. Nachr. {\bf 309} (1988) Nr. 4, pages 307 - 310;  
  Author's address that time:  
Zentralinstitut f\"ur  Astrophysik der AdW der DDR, 
1591 Potsdam, R.-Luxemburg-Str. 17a.}}

\end{document}